\documentclass[11pt,notitlepage]{isasrep}
\usepackage[dvips]{graphicx}
\usepackage[small,hang]{caption2}
\begin{document} 
\setcounter{page}{51}


\title{Mid-infrared Studies of the AGB Star Populations \\
      in the Local Group Galaxies : the Magellanic Clouds}
                        

\markboth{Ku\v{c}inskas et~al.}{AGB Star Populations in the Local Group}


\author{
     Arunas {\sc Ku\v{c}inskas}\footnotemark[1]$\;$,
     Vladas {\sc Vansevi\v{c}ius}\footnotemark[2]$\;$,
       Marc {\sc Sauvage}\footnotemark[3]$\;$, and          
     Toshihiko {\sc Tanab\'{e}}\footnotemark[4]$\;$ 
} 


\date{(November 1, 2000)} 

\maketitle  


\renewcommand{\thefootnote}{\fnsymbol{footnote}} 

\footnotetext[1]{National Astronomical Observatory, Tokyo, 181-8588, Japan; 
		arunaskc@cc.nao.ac.jp\\
            Institute of Theoretical Physics and Astronomy, Go\v {s}tauto 12, 
            Vilnius 2600, Lithuania}
\footnotetext[2]{Institute of Physics, Go\v {s}tauto 12, Vilnius 2600, Lithuania}
\footnotetext[3]{CEA/DSM/DAPNIA/Service d'Astrophys, C. E. Saclay, F-91191 
      Gif-sur-Yvette Cedex, France}
\footnotetext[4]{Institute of Astronomy, School of Science, The University of Tokyo, 
      Tokyo, 181-0015, Japan}

\renewcommand{\thefootnote}{\arabic{footnote}}   
\setcounter{footnote}{0}                   
\renewcommand{\baselinestretch}{1}         


\begin{abstract}
We present a brief summary of the ISO survey of AGB stars in 8 populous intermediate 
age clusters of the Magellanic Clouds. Totally, more than 100 AGB and RGB stars were 
observed at 4.5, 6.7 and 12 $\mu$m. We introduce a new method to estimate effective 
temperatures of oxygen-rich AGB stars, and derive precise  $M_{\rm bol}$ and 
$T_{\rm eff}$ for a large population of AGB objects in both LMC and SMC. The derived stellar 
parameters are used to construct observed HR diagrams which are employed further to 
estimate cluster ages using the isochrone fitting. We show that proposed method gives 
a powerful tool to study AGB star populations within the Local Group galaxies, especially, 
when backed up with the observations from forthcoming mid-IR astronomical space facilities 
(ASTRO-F, SIRTF) and large ground-based telescopes (SUBARU, VLT, etc.).
\end{abstract}


\section{INTRODUCTION}

Asymptotic giant branch (AGB) stars, as being one of the brightest stellar populations, 
can be effectively used to trace star formation histories within the Local Group galaxies. 
However, due to the sensitivity of presently available mid-infrared and far-infrared 
detectors, such studies so far were mostly confined to the Magellanic Clouds (MC). Indeed, 
populous intermediate age clusters in both the Large and Small Magellanic Clouds (LMC and SMC 
hereafter) represent an ideal laboratory for studying stellar evolution. They span a range 
of ages and metallicities, which makes them suitable to trace the AGB evolution of stars 
with different masses and in different evolutionary stages.

In this work we present a brief summary of ISOCAM survey of AGB stars in the populous 
intermediate age star clusters in the LMC and SMC. Although several obscured AGB stars 
were discovered during the course of our survey (Tanab\'e et~al. 1998, 1999), in 
this work we concentrate on non-obscured AGB stars. Since the radiation of these objects 
is not modified by the surrounding circumstellar shell, their spectral energy 
distributions (SEDs) can be effectively modeled using conventional model atmospheres, 
which allows a straightforward determination of $T_{\rm eff}$ and $M_{\rm bol}$. 

Although the AGB stars are ideal candles to trace stellar ages on extragalactic distance 
scales, an efficient method to derive ages of stellar populations using AGB stars is not yet 
available. We show that a precise knowledge of the basic stellar parameters of the AGB objects 
may provide a clue for the solution of this problem too.

\section{ISOCAM OBSERVATIONS}

The populous intermediate age clusters in the LMC and SMC were observed in the raster 
imaging mode with the ISOCAM (\cite{Cesarsky}) on board the ISO satellite (Kessler et~al. 1996).
Observations were made using the broad-band CAM filters LW1, LW2 and LW10 (corresponding 
to the effective wavelengths of 4.5, 6.7 and 12 $\mu$m, respectively), with a pixel field 
of view (PFOV) of 3$^{\prime\prime}$. The raster mode was typically 5 $\times$ 5 frames, with the raster 
step size equal to 8 pixels (24$^{\prime\prime}$). The fundamental integration time was 
set to $t_{\rm int}=2.1$ sec, with a total number of 15 to 20 exposures per single raster position. 

\begin{figure}[ht] 
   \begin{center}
   \resizebox{0.75\hsize}{!}{\includegraphics*[45,185][554,693]{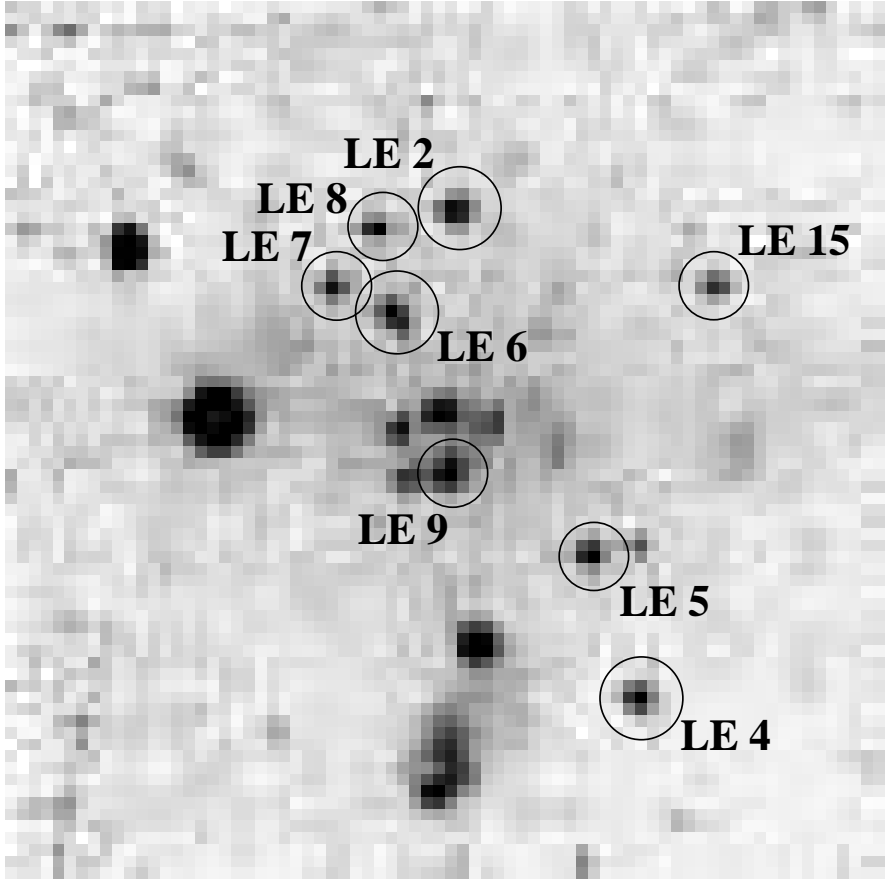}}
   \caption{ISOCAM image of the populous intermediate age cluster NGC 1783 in 
   the Large Magellanic Cloud at 6.7 $\mu$m. Marked are objects with our effective 
   temperature determinations (LE numbers are from Lloyd Evans 1980).}
   \label{fig1}
   \end{center}
\end{figure}

ISOCAM data were reduced using the CAM Interactive Analysis software (CIA version 3)
\footnote{The ISOCAM data presented in this paper was analyzed using ``CIA", a joint 
development by the ESA Astrophysics Division and the ISOCAM Consortium led by the ISOCAM 
PI, C. Cesarsky, Direction des Sciences de la Mati\`{e}re, C.E.A., France.}, and photometry 
was performed with the IRAF DAOPHOT package. The absolute photometric uncertainty of our 
measurements is mainly due to the ISOCAM calibration uncertainties and the correction for 
the memory effect (transient) of the ISOCAM, which is estimated to be less than 20\% 
(\cite{Biviano}). A typical ISOCAM image of one of the sample clusters (NGC 1783) is 
given in Figure~\ref{fig1}.

The obtained ISOCAM fluxes of the AGB stars, combined with the optical and near-infrared 
photometry available  from the literature, were used to construct the observed SEDs, which 
were further used to derive $T_{\rm eff}$ and $M_{\rm bol}$ of the AGB stars without circumstellar 
shells.

\section{$T_{\rm eff}$ AND $M_{\rm bol}$ OF THE STUDIED AGB STARS}

Though there are numerous methods to derive effective temperatures of the AGB stars, 
none of them works well in cases when limited information about the spectral energy 
distribution of a particular star is available (e.g. when only the  broad-band 
photometric fluxes are available). Various calibrations of photometric colors vs. effective
temprature have 
been used extensively in such cases (\cite{Bessell}; \cite{Feast}; \cite{Montegriffo}), 
though the accuracy of the derived $T_{\rm eff}$ is not better than $\sim200$ K. 
      
Since color indices measure only a portion of the observed SED, we suggest 
to use the entire SED as an indicator of the effective temperature. One of the possible 
ways is to employ synthetic SEDs of AGB stars, which would yield $T_{\rm eff}$ as a fitting 
parameter. In our analysis we have fitted the observed SEDs with the theoretical fluxes  
calculated from synthetic spectra of Fluks et~al. (1994). Though the synthetic spectra of 
Fluks et~al. (1994) are of solar metallicity, our analysis shows, that the influence of metallicity 
(and gravity) on the derived $T_{\rm eff}$ is minor (Ku\v{c}inskas et~al. in preperation).
Therefore, this offers a flexible
tool for estimating effective temperatures of AGB stars in stellar 
populations where the metallicity is not known {\it a priori}. It should be stressed,  
however, that this method would not be applicable in case of AGB stars with circumstellar 
shells.

\begin{figure}[ht] 
   \begin{center}
   \resizebox{0.75\hsize}{!}{\includegraphics*[50,80][550,780]{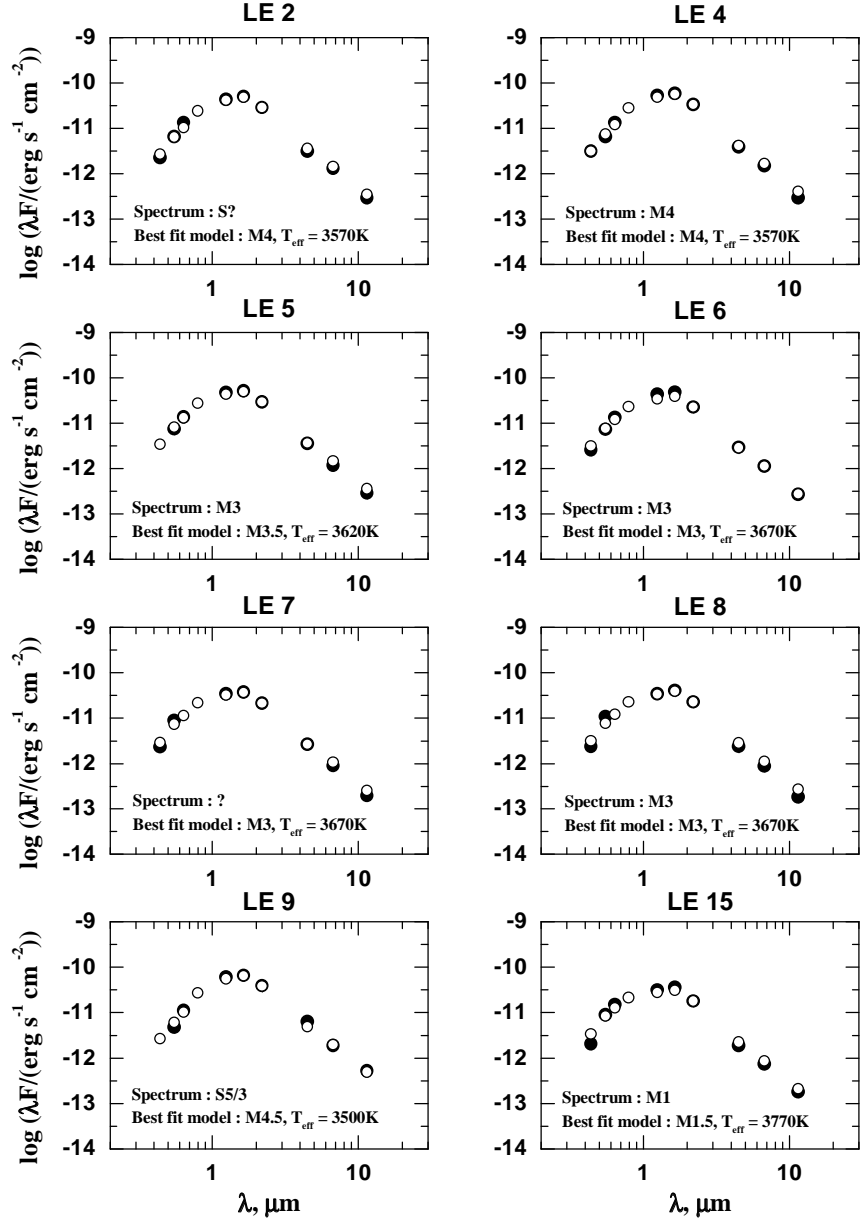}}
   \caption{Fits of the observed SEDs (solid circles) of oxygen-rich AGB stars in
   NGC 1783 with the synthetic SEDs (open circles). LE numbers are from Lloyd Evans 
   (1980). Indicated are spectral classes extracted from the literature (\cite{Bessell};
   \cite{Lloyd Evans 84}; \cite{Frogel}) and those derived in our work.}
   \label{fig2}
   \end{center}
\end{figure}

As an example, the observed SEDs and fitted synthetic SEDs are shown for the AGB stars  
in the LMC clusters NGC 1783 (Figure~\ref{fig2}). Typically, the accuracy 
of such fits is better than $\pm50$K. This therefore gives a possibility to obtain precise 
effective temperature estimates of the AGB stars even if only the broad-band photometric fluxes 
are available.

Absolute bolometric magnitudes, $M_{\rm bol}$, of the studied AGB objects were obtained by 
flux integration in the observed SEDs. To obtain the total integrated flux, the observed 
SEDs were linearly interpolated between the two adjacent photometric fluxes, whereas the 
blackbody extrapolation was used for both the short- and long-wavelength tails of the 
SEDs. The apparent bolometric magnitudes were calculated then using the elementary formula :

\begin{equation}
m_{\rm bol} = - 2.5 \log (F_{\rm tot}) + ZP
\end{equation}

where $F_{\rm tot}$ is the total integrated flux, and the zero point is taken to be $ZP= - 11.478$ 
(\cite{Montegriffo}). The apparent bolometric magnitudes were then converted into the 
absolute bolometric magnitudes using the distance moduli of 18.4 and 18.9 for the LMC 
and SMC, respectively.

It should be stressed, that the knowledge of the mid-infrared fluxes of AGB stars is 
important in several aspects when deriving $T_{\rm eff}$ and $M_{\rm bol}$ :

      - mid-infrared fluxes constrain the total amount of energy emitted by the star and 
      thus influences on the accuracy in determining $M_{\rm bol}$ (the improvement can reach 
      up to 10\% with the respect to $M_{\rm bol}$ derived from the SEDs constructed of optical 
      and near-infrared observations only, where the red-tail of the SED is extrapolated 
      as a blackbody); 

      - mid-infrared observations can reveal the excess flux due to the circumstellar dust, 
      thus resulting in an easy discrimination between the AGB stars with and without 
      circumstellar shells;

      - mid-infrared data constrain fits of theoretical SEDs to the observed SEDs in the 
      long-wavelength tail, and thus allow to derive $T_{\rm eff}$ more precisely.

\section{AGES OF THE POPULOUS STAR CLUSTERS IN THE LMC AND SMC}

\begin{figure}[ht] 
   \begin{center}
   \resizebox{0.70\hsize}{!}{\includegraphics*[60,160][520,730]{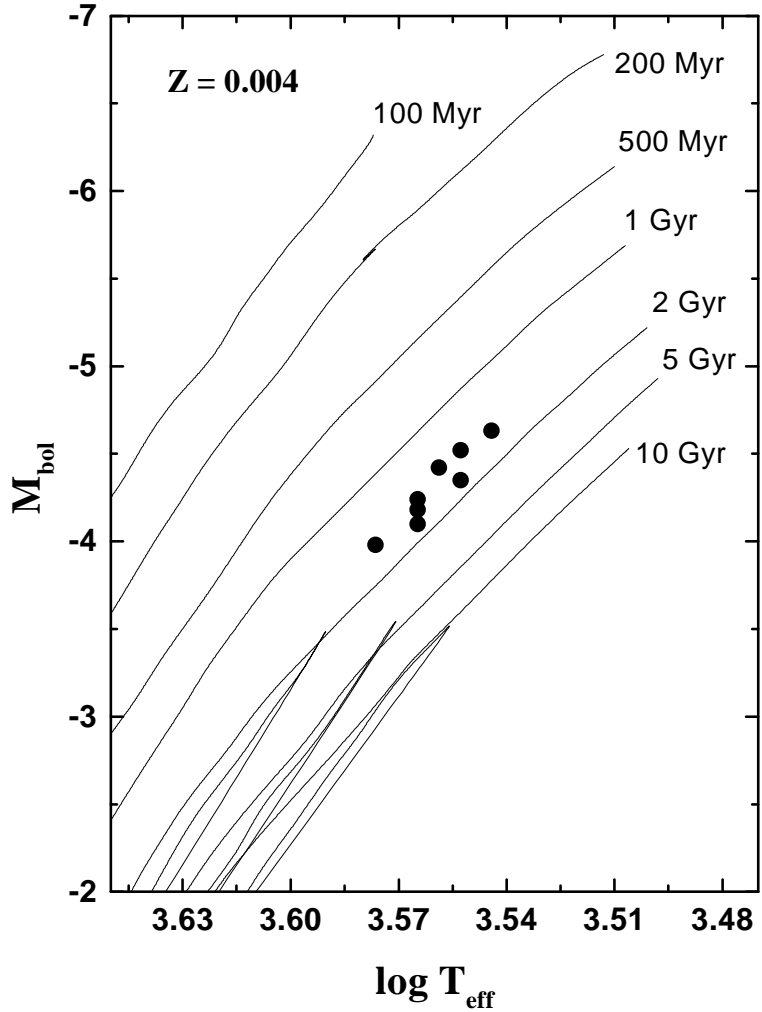}}
   \caption{HR diagram of the oxygen-rich AGB stars with synthetic SED fits in NGC 
   1783. Isochrones are from Bertelli et al. (1994) for the metallicity of $Z=0.004$ 
   (corresponding to the average metallicity from the determinations of Cohen (1982),
   Bica et~al. (1986) and de Freitas Pacheco et~al. (1998)).}
   \label{fig3}
   \end{center}
\end{figure}

The derived $T_{\rm eff}$ and $M_{\rm bol}$ of the non-obscured AGB stars were used to construct 
the observed HR diagrams, which were employed further to derive cluster ages using the 
isochrone fitting. We have used Bertelli et al. (1994) isochrones to work in $M_{\rm bol}$ 
versus $T_{\rm eff}$ plane. A typical example of the observed HR diagram and isochrone fitting 
is given in Figure~\ref{fig3} for the LMC cluster NGC 1783. The derived ages of all clusters 
in the sample and their comparison with the ages derived from the main sequence turn-off point 
(MSTO) fitting are given in Table~\ref{tab1}. There is an excellent agreement between 
the ages obtained using these two methods.

It should be stressed, however, that metallicity plays an important role in the age 
determination. For instance, in case of NGC 1783 metallicity is not known precisely 
and the existing estimates vary between [Fe/H]$=-0.45$ and [Fe/H]$=-0.90$ (\cite{Cohen}; 
\cite{Bica}; \cite{de Freitas Pacheco}). The ages resulting from the isochrones of the 
corresponding metallicities would be $0.8\pm0.2$ Gyr and $2.2\pm0.4$ Gyr, respectively. 
It therefore shows, that {\it ab initio} knowledge of the cluster metallicity is an 
essential ingredient for obtaining a precise estimate of the population age.

\begin{table}[htbp]
\begin{center}
\caption{Ages of the sample clusters.}
\label{tab1}
\begin{tabular}{ccc} \hline \hline
Cluster       & MSTO estimate$^{1}$ (Gyr)    & This work$^{2}$ (Gyr)   \\ \hline
SMC         &                              &                         \\
Kron 3        & $8\pm2$                      & $9\pm3$                 \\
NGC 152       & $0.8$                        & $1.0\pm0.3$             \\
NGC 419       & $1.2\pm0.5$                  & $1.4\pm0.2$             \\
LMC         &                              &                         \\
NGC 1783      & $0.9\pm0.4$                  & $0.8\pm0.2$             \\
NGC 1846      & $-$                          & $1.3\pm0.3$             \\
NGC 1978      & $2.0\pm0.2$                  & $1.5\pm0.5$             \\
NGC 1987      & $-$                          & $4.0\pm2.0$             \\
NGC 2121      & 4.5                          & $6.0\pm2.0$             \\ \hline \\
\end{tabular} 
\end{center}
$^{1}$ All ages in this column are from MSTO fits except for NGC 152 where 
      it comes from the CMD fit without the MSTO region. References for the age estimates 
      of individual clusters are as follows : Kron 3 - Alcaino et~al. (1996); NGC 152 -
      Hodge (1981); NGC 419 - Durand et~al. (1984); NGC 1783 - Mould et~al. (1989);
      NGC 1978 - Bomans et~al. (1995); NGC 2121 - Sarajedini (1998).

$^{2}$ Ages listed in this column were derived employing cluster metallicities 
      used to obtain MSTO age estimates listed in Column 2. In case of  NGC 1846 and 
      NGC 1987 isochrones corresponding to the metallicity of $Z=0.004$ were used.
\end{table}

\section{SUMMARY AND FUTURE PROSPECTS}

We suggest an alternative way for obtaining precise effective temperatures of 
oxygen-rich AGB stars without circumstellar shells. The only information needed to 
use this method is the availability of photometric broad-band fluxes. The observed 
SEDs should span as large wavelength interval as possible (optical through mid-infrared), 
which makes the mid-infrared observations highly desirable.

The obtained fundamental parameters of non-obscured oxygen-rich AGB stars 
($T_{\rm eff}$ and $M_{\rm bol}$) can be used further to derive population ages employing 
the usual procedure of isochrone fitting. This gives a possibility to study star 
formation histories in a wide range of ages and metallicities.

The present ISOCAM study allowed to detect non-obscured AGB and RGB stars in the LMC 
and the SMC down to $M_{\rm bol}\sim-2.7$ ($L_{\star}\sim1000 L_{\odot}$). The forthcoming 
infrared space missions (ASTRO-F, SIRTF etc.) and the ground-based instrumentation for 
large telescopes (SUBARU, VLT etc.) will allow to detect and study  extragalactic 
AGB stars in most of the Local Group galaxies (\cite{Zijlstra}; \cite{Kaufl}). 
Therefore, such observations combined with the proposed method to estimate $T_{\rm eff}$ 
and age offer a powerful tool to study evolution of stellar populations in diverse 
astrophysical environments.


\section*{ACKNOWLEDGMENT}  

This research was supported in part by grant-in-aids for Scientific research (C) 
and for International Scientific research (Joint Research) from the Ministry of 
Education, Science, Sports and Culture of Japan.


\end{document}